\documentclass[fleqn,usenatbib]{mnras}

\usepackage[T1]{fontenc}
\usepackage{ae,aecompl}

\usepackage{graphicx}	
\usepackage{amsmath}	
\usepackage{amssymb}	
\usepackage[normalem]{ulem} 

\newcommand{\D}{\mathrm{d}}
\newcommand{\Ms}{{\ensuremath{\mathrm{M}_{\sun}}}}
\newcommand{\Zs}{{\ensuremath{\mathrm{Z}_{\sun}}}}

\title[Minimal dilution]{A Minimum Dilution Scenario for Supernovae and Consequences for Extremely Metal-Poor Stars}

\author[M. Magg et al.]{Mattis Magg$^{1,2}$\thanks{E-mail: mattis.magg@stud.uni-heidelberg.de}, Thomas Nordlander$^{3,4}$,  Simon C. O. Glover$^{1}$, Camilla J. Hansen$^5$,
\newauthor  Miho Ishigaki$^{6,7}$, Alexander Heger$^{4,8-11}$, Ralf S. Klessen$^{1,12}$, Chiaki Kobayashi$^{13}$
\newauthor and Ken'ichi Nomoto$^{14}$\\
$^{1}$Universit\"at Heidelberg, Zentrum f\"ur Astronomie, Institut f\"ur Theoretische Astrophysik, D-69120 Heidelberg, Germany\\
$^{2}$International Max Planck Research School for Astronomy and Cosmic Physics at the University of Heidelberg (IMPRS-HD)\\
${^3}$Research School of Astronomy and Astrophysics, Australian National University, Canberra, ACT 2611, Australia\\
${^4}$ARC Centre of Excellence for All Sky Astrophysics in 3 Dimensions (ASTRO 3D), Australia\\
$^{5}$Max Planck Institute for Astronomy, K\"onigstuhl 17, D-69117 Heidelberg, Germany\\
$^{6}$Tohoku University, Aoba, Sendai 980-8578, Japan\\
$^{7}$National Astronomical Observatory, 2-21-1 Osawa, Mitaka, Tokyo 181-8588, Japan \\
$^{8}$School of Physics and Astronomy, 19 Rainforest Walk, Monash University, VIC 3800, Australia\\
$^{9}$Australian Research Council Centre of Excellence for Gravitational Wave Discovery (OzGrav), Clayton, VIC 3800, Australia\\
$^{10}$Joint Institute for Nuclear Astrophysics, 1 Cyclotron Laboratory, National Superconducting Cyclotron Laboratory,\\
\phantom{$^7$}Michigan State University, East Lansing, MI 48824-1321, USA\\
$^{11}$Tsung-Dao Lee Institute, Shanghai 200240, China\\
$^{12}$Universit\"{a}t Heidelberg, Interdiszipli\"{a}res Zentrum f\"{u}r Wissenschaftliches Rechnen, D-69120 Heidelberg, Germany\\
$^{13}$Centre for Astrophysics Research, University of Hertfordshire, College Lane, Hatfield AL10 9AB, UK\\
$^{14}$Kavli Institute for the Physics and Mathematics of the Universe (WPI), The University of Tokyo, Kashiwa, Chiba 277-8583, Japan
}

\date{Accepted XXX. Received YYY; in original form ZZZ}

\pubyear{2020}

\begin{document}
\label{firstpage}
\pagerange{\pageref{firstpage}--\pageref{lastpage}}
\maketitle

\begin{abstract}
To date no metal-free stars have been identified by direct observations. The most common method of constraining their properties is searching the spectra of the most metal-poor stars for the chemical elements created in the first stars and their supernova. In this approach, modelled supernova yields are compared to the observed abundance patterns in extremely metal-poor stars.  The method typically only uses the abundance ratios, i.e., the yields are diluted to the observed level.  Following the usual assumption of spherical symmetry we compute a simple lower limit of the mass a supernova can mix with and find that it is consistent with all published simulations of early chemical enrichment  in the interstellar medium. For three different cases, we demonstrate that this dilution limit can change the conclusions from the abundance fitting. There is a large discrepancy between the dilution found in simulations of SN explosions in minihaloes and the dilution assumed in many abundance fits. Limiting the dilution can significantly alter the likelihood of which supernovae are possible progenitors of observed CEMP-no stars. In particular, some of the faint, very low-yield SNe, which have been suggested as models for the abundance pattern of SMSS0313-6708, cannot explain the measured metal abundances, as their predicted metal yields are too small by two orders of magnitude. Altogether, the new dilution model presented here emphasizes the need to better understand the mixing and dilution behaviour of aspherical SNe.
\end{abstract}




\section{INTRODUCTION}
The first stars, so-called Population~III (Pop~III) stars form in the absence of heavy elements in the early Universe. Due to the lack of metal cooling, they are expected to be drastically different from the stars found in our vicinity at the present day \citep{BrommReview, GloverReview, GreifReview}. Initially, Pop~III stars were thought to be very massive \citep[e.g.,][]{Bromm99, Omukai01}, but later it was found that their protostellar disks may fragment, leading to the formation of clusters of low-mass metal-free stars \citep{Greif11b, Clark11}. Whereas it is clear that Pop~III stars may form over a wide range of masses, until today simulations are unable to constrain well the metal-free initial mass functions (IMFs). The results depend significantly on the physics employed, the choice of numerical method, and the resolution \citep[see e.g.,][]{Hosokawa16, Stacy16, Susa19}.

There are so far no direct detections of metal-free stars. Pop~III stars are expected to form in high-redshift, relatively low-mass mini- and atomic-cooling haloes. Therefore, ``Pop~III galaxies'' are most likely not bright enough to be detected today \citep{Xu16, Hartwig16b, Visbal17}. Having no direct observations, there are several indirect methods that allow us to gain observational constraints on the IMF of Pop~III stars. Direct detection of supernovae (SNe) \citep{Hummel12, Hartwig18b, Rydberg20} or gravitational waves \citep{Kinugawa14, Kinugawa16, Hartwig16a} from the first stars are challenging, but may provide constraints on the high-mass end of the Pop~III IMF in the coming decade. The 21\,cm absorption feature, as reported by the EDGES experiment \citep{EDGES18}, can constrain the timing of the first star formation and the star formation efficiency, but it is not very sensitive to the assumed IMF \citep{Schauer19}.

There are two remaining methods to constrain the pristine IMF that are feasible at present. Both are related to the observations of ancient metal-poor stars in the Milky Way and its satellites. The first one is constraining the low-mass end of the IMF with the current non-detection of metal-free stars \citep{Salvadori07, Hartwig15b, Ishiyama16, Magg18, Magg19}. The second method, which is our focus here, is comparing the abundance patterns observed in metal-poor stars to simulated SN yields in Pop~III stars. It was found that the most metal-poor stars, often called extremely metal poor (EMP) stars with an iron abundance\footnote{For elemental abundances, we use the notation \\$\mbox{[X/H]} = \log_{10}(N_\mathrm{X}/N_\mathrm{H})-\log_{10} (N_{\mathrm{X},\sun} /N_{\mathrm{H},\sun})$ where $N_{\mathrm{X}}$ and $N_{\mathrm{H}}$ are the fractional abundances of any element X and hydrogen, and $N_{\mathrm{X},\sun}$ and $N_{\mathrm{H},\sun}$ are the corresponding solar abundances.}  of less than $\mbox{[Fe/H]} =-3$ are surprisingly rich in carbon \citep{Beers2005, Frebel15}. A particularly interesting subgroup of stars are the carbon-enhanced extremely metal-poor (CEMP) stars and among these the CEMP-no stars: those with iron abundances below  $\mbox{[Fe/H]}=-3$, an excess of carbon relative to iron of more than  $\mbox{[C/Fe]}=1$ and no enhancement in neutron-capture elements, i.e.,  $\mbox{[Ba/Fe]}<0$\ \citep[e.g.,][]{Frebel05, Keller14, Aguado18, Nordlander19}. The origin of the elemental abundance patterns in these stars is one of the key questions of early chemical enrichment. 

\citet{Umeda2003} proposed that the abundance pattern of CEMP-no stars is the fingerprint of  so-called faint SNe. These are Pop~III SNe with relatively large mixing-and-fallback at the core-collapse explosions\footnote{The explosion energy and progenitor star mass are not necessary larger than those for Pop~II SNe \citep{kob14}.}, and eject much of their outer layers, containing carbon and other light elements, whereas most of the inner shells, containing in particular iron, fall back onto the compact remnant. Thus, when the first metal-enriched stars form from gas enriched by one of these SNe, they form with a very small iron abundance but a much higher carbon abundance. Notably, these SNe do not produce particularly large absolute amounts of carbon compared to more conventional core-collapse SNe; rather, they yield high [C/Fe] ratios because they produce unusually small amounts of iron.

Subsequently, it has become common practice to use SN models to infer the properties of the primordial progenitors of the most metal-poor observed stars, both for individual stars \citep[e.g.,][]{HegerWoosley2010, Hansen11, Nomoto13, Ishigaki14, Bessel15, Placco16} and for large samples \citep{Cayrel04, Fraser17, Ishigaki18}. Additionally, constraints on the primordial IMF can be inferred from bulk properties of metal-poor stars with semi-analytical models \citep{deBennassuti17, Hartwig18b, Tarumi20b}. For these purposes, libraries of SN yields have been computed \citep{HegerWoosley2010, Nomoto13, Ishigaki18}.  These yields typically depend on the stellar mass of the exploding star, the explosion energy and one or a few parameters that quantify the mixing-and-fallback process, which cannot be simulated self-consistently in the one-dimensional SN simulations \citep[e.g.,][]{Chen17, Chan20}.  Comparing the resultant Fe mass to the observed [Fe/H], dilution has been discussed in \citet{tom07} and \citet{kob11}.

In order to compare modelled and observed abundance pattern, one further step is required: the SN yields need to be physically diluted with metal-free gas to match the absolute metallicity of the observed star. Usually this dilution is treated as a free parameter and chosen to optimize the quality of fit. Freely adjusting the dilution factor essentially makes the fit independent of the absolute abundances and only considers the ratios of the abundances to each other. Then observed and modelled abundance pattern are compared and the well-fitting ones are interpreted as likely progenitor. For example, the \textsc{starfit}\footnote{\url{http://starfit.org}} \citep{HegerWoosley2010, Fraser17} pipeline can be used for such an analysis.

In this study, we argue that this approach has do be amended because the amount of ambient gas into which the metals from a Pop~III supernova are mixed cannot be assumed to be arbitrarily small. We derive a simple analytical model for the lower limit of the mass a SN remnant has to mix with before it can recollapse. We find that this limit is consistent with the results from 3D hydrodynamical simulations. In many cases, there are large differences between the halo-scale mixing found in hydrodynamical simulations and the mixing implicitly assumed by fitting abundance ratios with arbitrary dilution. We show how the dilution limit can be applied in abundance fitting methods. Finally, we investigate examples of the impact this dilution limit has on the conclusions drawn from fitting observed abundances.

\section{The minimum mixing mass}
\subsection{Analytic estimate}
 As outlined before, abundance fitting usually employs the observed ratios of abundances of certain metals and compares those to the ratios found in theoretical SN models. Of particular importance is, e.g., the $[\mathrm{C}/\mathrm{Fe}]$ ratio. This method, however, typically neglects the absolute abundances (i.e., [Fe/H] or [C/H]) and treats them as an arbitrary normalization factor. Conceptually, this normalization can be achieved by diluting the SN yields with the correct amount of metal-free gas. As in published work usually only single SNe are fitted to observed abundance patterns, we only consider single, isolated SNe in this work.
 
We consider SN explosions as well as their subsequent expansion into the ambient medium and the corresponding mixing processes in spherical symmetry.  Simulations carried out in two \citep{Tominaga09} and three \citep{Chan20} dimensions, however, show that Pop~III SNe can be strongly aspherical.  In this context, we note that even when considering anisotropic SNe, the observed abundances in most published studies are compared to angle-integrated yields. This means that the problem considered is effectively spherically symmetric, as the angular average implies that different elements ejected in different directions become well mixed before the second generation stars form. An exception to this may be, if the abundances are distributed more spherically than the energy input, such as seen in some of the models in \citet{Tominaga09}. A critical analysis of the validity of this approximation is one of the primary motives for the study presented here.  We argue that properly accounting for the asymmetries expected in Pop~III SNe requires both detailed three dimensional explosion models as well as high-resolution simulations of the expansion of the resulting anisotropic shock wave into an inhomogeneous ambient medium that are able to adequately follow the chemical mixing process.
 
Since there is no analytic model for such a small-scale inhomogeneous mixing, however, we  follow the bulk of the existing literature and approximate the SN as a spherical explosion inside a homogeneous ambient medium. As SNe are very energetic events, a large amount of gas is required to confine the metals and thus not all dilution masses are physically plausible. The lowest limit for this mass is the mass enclosed in the final radius of the SN remnant. Analytical solutions to spherical blast waves of SNe can be derived under a variety of assumptions \cite[ e.g.,][]{Ostriker88}, with the expansion of the remnant stalling at the end of the momentum-driven snowplough phase. In this phase, the expansion velocity reaches the speed of sound in the ambient medium. As shocks cannot be subsonic, the shock wave transforms into a sound wave and dissipates. This occurs at the fade-away radius $R_\mathrm{fade}$ which is
 \begin{equation}
  R_\mathrm{fade} \approx 2.07\times 10^{20}\,\mathrm{cm}\ E_{51}^{0.32} n_0^{-0.37} \left(\frac{c_\mathrm{s}}{10\,\mathrm{km}\,\mathrm{s}^{-1}}\right)^{-2/5},
  \label{eq:R_fade}
 \end{equation} 
 where $n_0$ is the nucleon number density of the ambient medium in units of cm$^{-3}$, $E_{51}$ is the explosion energy in units of $10^{51}\,\mathrm{erg}$ and $c_\mathrm{s}$ is the ambient medium speed of sound \citep[e.g.,][]{Draine}. We assume the ambient medium is ionized, i.e., that it has a speed of sound of $c_\mathrm{s}=18\,\mathrm{km}\,\mathrm{s}^{-1}$, for a metal-free H\textsc{ii} region \citep[see e.g.,][]{abel07}. In case the medium is actually neutral, the speed of sound would be lower and the stalling radius larger. Thus, this is a conservative assumption. In the homogeneous mixing case, the minimum mass with which the ejecta are mixed is the mass that is enclosed in the stalling radius, i.e.,
 \begin{equation}
  M_\mathrm{dil, min} = \frac{4}{3}\pi n_0 \mu m_{\rm H} R_\mathrm{fade}^3 = 1.9\times 10^4\,\Ms\,E_{51}^{0.96}\,n_0^{-0.11},
  \label{eq:M_dil}
 \end{equation}
where $m_{\rm H}$ is the mass of a hydrogen nucleus and where we assumed a mean molecular weight of $\mu=1.22$. The fade-away radius used here is for gas cooling rates of solar metallicity gas. However, the smaller amount of metals in the case considered here would only decrease the cooling rates and therefore increase the total mixing mass. As we aim at computing a lower limit for the mixing mass, the reduced cooling can be neglected. By definition the SN remnant expands faster than the speed of sound in the ionized medium. As we consider haloes below the atomic cooling limit the escape velocity from the haloes is much smaller than this speed of sound. Therefore SN remnants expand much faster than the escape velocity and the effect of gravity can be neglected.
 
 This result is very similar to the one obtained through numerical simulations by \citet{Thornton98}. While it has been widely used in the discussion of stellar feedback, it is often neglected when fitting abundance patterns of individual stars. For example, \citet{Tominaga14} note that the minimum dilution mass obtained by \citet{Thornton98} is not a binding limit, as metal mixing is highly inhomogeneous \citep{Ritter12}. We will later see that our derived limit holds even in cases of inhomogeneous mixing.
 
 We assume an ambient density of $n_0 = 1\,\mathrm{cm}^{-3}$, which should be the typical case for the ionized regions around massive Pop~III stars \citep{Whalen04}. We note that the density dependence of the minimum mixing mass (Eq. \ref{eq:M_dil}) is very weak, so it would need to be higher by several orders of magnitude to affect our conclusions. If the density is this much higher than the assumed value, the free-fall time of the ambient gas is smaller than the life-time of the star, and thus it should form stars already before the SN explodes or while the remnant expands. Furthermore, simulations show that it is difficult to mix metals into gas that is already very dense when the SN explodes \citep{Ritter16, Chiaki18}.
 
 Under the assumptions outlined above, the dilution mass is lower limit for two main reasons: 
 \begin{enumerate}
  \item We assume a homogeneous medium. If the medium is not homogeneous the denser gas will be less enriched but form stars first. This effect is discussed further below.
  \item We assume no further mixing. Realistically further mixing with additional pristine gas should occur during recollapse, rather than the stalled SN remnant monolithically collapsing back on itself. This effect would further increase the dilution mass.
 \end{enumerate}
We note that we assume all SNe are able to produce second generation stars. Very energetic explosions may actually disrupt their host haloes, which suppresses or delays second generation star formation \citep{Whalen08b}. This effect is difficult to quantify without hydrodynamical simulations in cosmological context, and is therefore neglected here.
 
 \subsection{Consistency with simulations}

To see whether sub-galactic-scale inhomogeneous mixing can lead to higher metallicities than predicted by the minimum dilution we will compare it to the dilution found in all suitable published simulations of inhomogeneous mixing and the formation of second generation stars which we are aware of. For comparison with our limit, we use an ambient density of $n_0 = 1\,\mathrm{cm}^{-3}$ in all cases but take the explosion energies used in the simulations to compute the minimum mixing.  Simulations are included provided that they
\begin{itemize}
    \item are three dimensional hydrodynamical simulations of the expansion of Pop~III SN remnants into their ambient medium,
    \item are set up to and have sufficient resolution to model individual, isolated, Pop~III SNe, not combined populations,
    \item follow the enriched gas until it re-collapses, and
    \item provides the output needed for our comparison.
\end{itemize}
This implies that we do not discuss the results from \citet{Greif07} and \citet{Chen15} because the re-collapse of the enriched gas is not modelled. Larger-scale simulations, such as the ones from \citet{Wise12}, \citet{FiBY1}, or \citet{Tarumi20a}, are not considered, because they do not follow individual isolated SNe. The simulations of \citet{Whalen08b} are not included here because they are one-dimensional. Nevertheless, we note that their metal-enriched gas masses are consistent with our upper limit in most cases.  Only in one of their models is the enriched gas mass they find smaller than our prediction in Eq. \eqref{eq:M_dil}. In this case, the star completely fails to create an ionized region, and the ability to model an off-centre re-collapse would be crucial to make accurate predictions for the metallicity of the second generations star.

We begin with the dilution found in \citet{Ritter12, Ritter15, Ritter16}.\footnote{We only consider the 1\textsc{sn} model from \citet{Ritter15} as the 7\textsc{sn} model deals with enrichment by multiple SNe, which is not the topic of our analysis.} In all three simulations the SNe considered are core collapse (CC) SNe with $E_{51}=1$. They eject $M_\mathrm{met}=4\,\Ms$ of metals in \citet{Ritter12, Ritter15} and  $M_\mathrm{met}=6\,\Ms$ in \citet{Ritter16}. Thus, according to Eq. \eqref{eq:M_dil} the maximum final metallicity we should expect is
 \begin{equation}
  Z_\mathrm{max} = \frac{M_\mathrm{met}}{M_\mathrm{dil, min}} \approx 10^{-3.6} \approx 10^{-1.7}\,\Zs
  \label{eq:ZMass}
 \end{equation}
where $\Zs=0.0142$ is the solar metallicity \citep{Asplund09}. While the mixing is highly inhomogeneous, and orders of magnitude of spread in metallicity can be seen, the newly collapsing cores always show metallicities below this value. All simulations also contain gas at higher metallicities than predicted by the minimum dilution. While from \citet{Ritter12} it is unclear in which phase this gas is contained, in \citet{Ritter15, Ritter16} only some of the very diffuse gas has metallicites above the dilution limit.
 
  \citet{Chiaki18} and \citet{Chiaki19} model the inhomogeneous mixing occurring after the SNe of 7 different stars with masses between 13\,\Ms\ and 200\,\Ms. Some of these SNe are simulated in several different halos. The simulations cover a wide range of different environments in which SNe can explode. For massive stars, halos are often completely photo-evaporated, whereas, for the smallest stars, the gas in the stellar birth-cloud remains dense throughout the lifetime of the star. The results show large variations between the mixing behavior and the metallicities of the second-generation stars. \citet{Chiaki18} distinguish between three separate enrichment channels:
 \begin{enumerate}
  \item Internal enrichment: in this case, the SN expands efficiently and the metals mix well with the surrounding gas before the halo collapses back on itself.
  \item External enrichment: the metals escape from the halo in which the SN explodes and mix with the gas in a different halo that has not formed stars yet. This type of enrichment is also found in \citet{bsmith15}.
  \item Inefficient internal enrichment: dense structures remain in the halo. When the SN explodes these structures are only enriched to very low metallicities and proceed to form stars with metallicities much lower than the average gas metallicity in the halo.
 \end{enumerate}
 None of these simulations, however, show the formation of second generation stars that violate our dilution limit.  According to Eq.~\eqref{eq:M_dil} the predicted maximum metallicity ranges between $10^{-2.6}\, \Zs < Z_\mathrm{max} < 10^{-1.6}\,\Zs$. All second generation stars in their simulations have metallicities in the range $10^{-6.3}\,\Zs<Z< 10^{-2.2}\,\Zs$. None of the stars violate our derived limit. The simulated second generation star that is closest to our computed upper limit is enriched by a 25\,\Ms\ CCSN that explodes in their halo ``MH1'', which is their smallest halo with a mass $M_\mathrm{vir}=3\times10^5\,\Ms$. The re-collapsing region has a metallicity of 40 per cent of our computed upper limit. In these simulations, there are several cases of stars with much lower metallicities than predicted by the minimum dilution model. These are the cases in which the surroundings of the SNe are the most dense and the mixing is the most inhomogeneous. The second generation stars form in clumps that already exist when the SNe explode and only the outer layers of these clumps are enriched with metals. Thus, the enrichment proceeds in what \citet{Chiaki18} label the ``inefficient internal enrichment'' channel. 
 
 \citet{Greif10} simulate the explosion of a single PISN with $E_{51}=10$ and $100\,\Ms$ of metal ejecta. According to our model the maximum metallicity in this extreme case should be below $Z=10^{-1.4}\,\Zs$. They find metallicities in the recollapsing galaxy that are around $Z=10^{-3}\,\Zs$. As \citet{Greif10} note, the average metallicities are initially much higher but they decrease to this low value during the recollapse of the halo, which takes around 300\,Myr.
 
 The simulations by \citet{Jeon14} include several SNe exploding in three different haloes. The authors provide information on the metallicity of recollapsing regions in three cases: a 15, 25 and 40\,\Ms\ star exploding in their ``halo1''. They all explode as $E_{51}=1$ CCSN and eject 5 per cent of their stellar mass as metals. According to our model, this should lead to metallicites of $Z<10^{-2.1}\,\Zs$. Their reported metallicities are all below $Z=10^{-3.5}\,\Zs$.
 
 \citet{bsmith15} highlight the external enrichment channel. Their SN is a $E_{51}=1$ CCSN which ejects $11.19\,\Ms$ of metals, leading us to predict a maximum metallicity of $Z_\mathrm{max} = 10^{-1.4}\,\Zs$. Only a very small fraction of gas is found at such high metllicities, and none of it is in the re-collapsing region. The metal-enriched star forming gas in this case has a metallicity of $Z=10^{-4.7}\,\Zs$.

\begin{figure}
  \includegraphics[width=\linewidth]{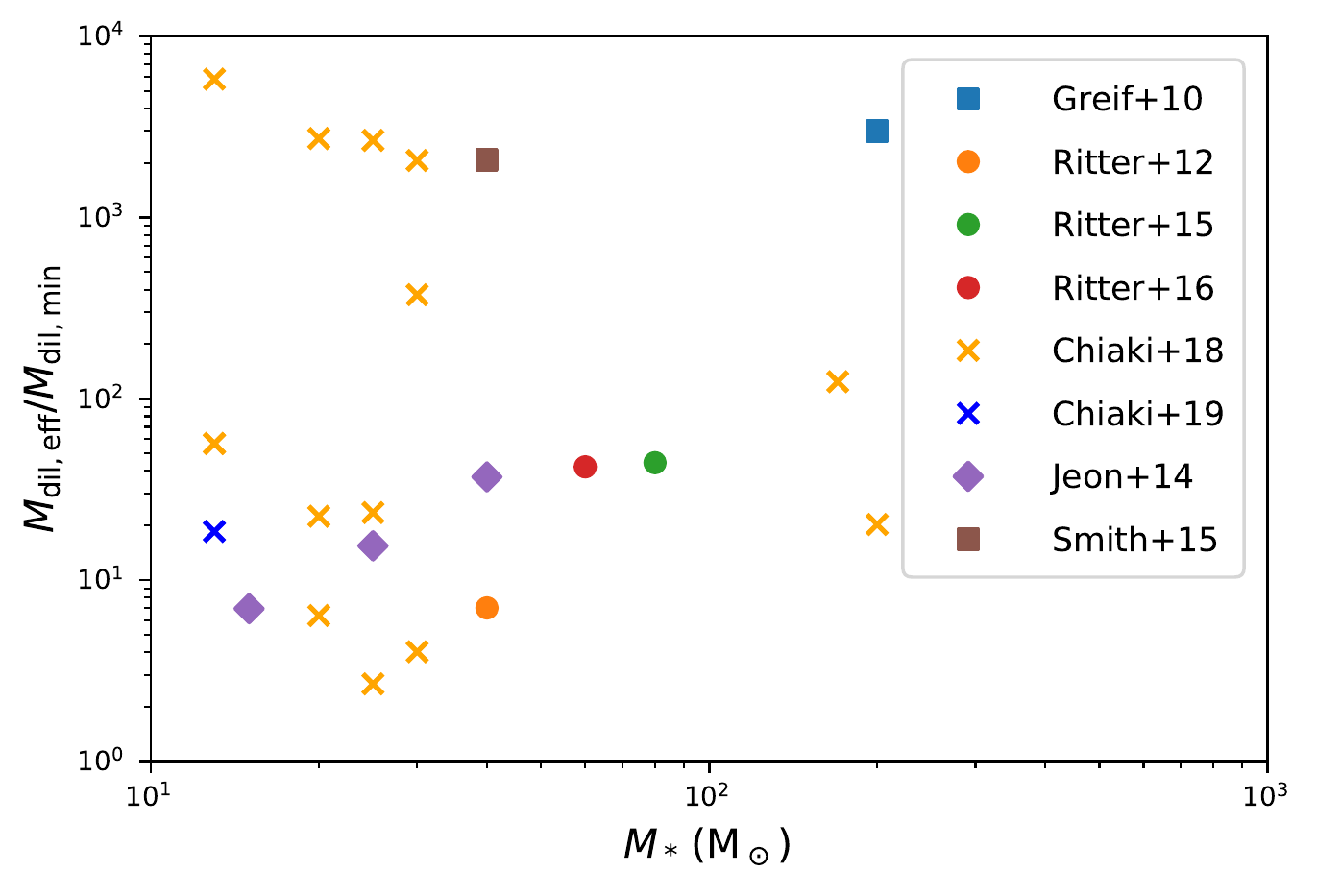}
  \caption{\label{fig:sim_comp} Comparison of the minimum dilution model to simulations of inhomogeneous metal mixing. We show the ratio of the effective dilution mass of the simulations and our estimate of the minimum dilution mass as a function of the stellar mass of the exploding star. The effective dilution mass is derived from the metallicity in the second generation stars or the likely sites of second generation star formation in the simulations. All simulations show a ratio above one, i.e. they are consistent with our predicted minimum.}
 \end{figure}
 
 We convert the metallicities found in the simulations back to an ``effective dilution mass'' with eq. \eqref{eq:ZMass} and summarize the simulations in Fig. \ref{fig:sim_comp}. None of the simulations of inhomogeneous mixing show inconsistencies with the minimum dilution mass derived from the spherically symmetric case. In some of the simulations, there is gas above the derived upper limit for the metallicity, but it tends to be diffuse and hot. This can be understood intuitively: as thermal energy and metals are ejected together, more metal-rich gas tends to be hotter. It is important to note that there is significant scatter in the simulation results: even for similar exploding stars the effective dilution mass can vary by many orders of magnitude. The cases with the largest effective dilution masses are usually external or inefficient internal enrichment. We conclude that, to the best of our knowledge and the current state of modelling, our estimate provides a useful limit on the mixing and dilution of metals even in the presence of inhomogeneous mixing.
 
 \subsection{Bayesian fitting}
 
 We will here briefly discuss how the derived limit on mixing can be implemented in abundance fitting codes. For this purpose we create an algorithm that fits observed abundances by comparing to them to the modelled SN yields from \citet{HegerWoosley2010}. The yields of the SNe generally depend on the progenitor mass ($M_\mathrm{prog}$), the explosion energy ($E_{51}$ in units of $10^{51}\,\mathrm{erg}$) as well as a mixing factor ($f_\mathrm{mix}$).  For matching observed and modelled abundances, we use the SN yields and analysis tools provided with \textsc{starfit} and supplement them with a generic Bayesian fitting approach. A general description of Bayesian parameter estimation can be found in \citet{BailerJones2017}. We first compute the likelihoods $L_i (x_i|M)$ that a model $M$, which predicts the abundances $y_i$, results in the observed abundances $x_i$,
 \begin{equation}
  L_i(x_i|M) = \exp\left(-\frac{(x_i-y_i)^2}{2\sigma_i^2}\right),
 \end{equation}
 where $i$ is any of the observed elements and $\sigma_i$ is the error of the observations. The normalization is left arbitrary for now. This likelihood calculation implicitly assumes that the errors follow a Gaussian distribution. While it is not clear whether this assumption is valid, it is commonly made when fitting SN models to observed abundances \citep[e.g.][]{HegerWoosley2010, Ishigaki18, Ezzeddine19}. Computing the modelled abundances $y_i$ requires, as discussed above, a usually arbitrary dilution mass $M_\mathrm{dil}$. If $M_i$ is the mass of element $i$, which has a mass-number of $\mu_i$, that is ejected by a SN, the model abundance is
\begin{equation}
 y_i = \log_{10} \left(\frac{M_i}{\mu_i\,X_\mathrm{H} M_\mathrm{dil}}\right) - \log_{10} \left(\frac{N_{i,\odot}}{N_{\mathrm{H},\odot}}\right),
\end{equation}
where $X_\mathrm{H}=0.754$ is the hydrogen abundance of primordial gas \citep{Planck2015}. The respective solar fractions of the element $i$ and of hydrogen are $N_{i,\odot}$ and $N_{\mathrm{H},\odot}$. We iteratively adjust the dilution mass for each model until we find the dilution that gives the maximum final likelihood according to Eq. \eqref{eq:comb_L}. The dilution mass is picked individually for each model, but within each model the same dilution mass is used for every element. This choice comes from our assumption that each element mixes in the same way. This assumption is commonly made for SN fitting, as without it the SN yields would not be representative of the elements found in the second generation stars.
 
 However, in many cases elements are not detected and only upper limits on their abundance can be derived. These upper limits need to be treated simultaneously with the detections. For this we assume that the upper limits are strict (i.e.\ the likelihoods are Heaviside step-functions $\Theta$) combined with a Gaussian error on where exactly this limit is. These assumptions lead to a likelihood $L_i(x_i|M)$ for an upper limit of $x_i$ in element $i$ of:
 \begin{equation}
 \begin{split}
  L_i(x_i|M) &= \int_{-\infty}^{\infty} \Theta(x_i-z_i) \exp\left(-\frac{(z_i-y_i)^2}{2\sigma_i^2}\right) \D z_i\\
  &=\int_{-\infty}^{x_i} \exp\left(-\frac{(z_i-y_i)^2}{2\sigma_i^2}\right) \D z_i\\
  &= \sqrt{\frac{\pi}{2}}\sigma_i\, \mathrm{erf} \left(\frac{x_i-y_i}{\sqrt{2}\sigma_i} \right),\\
 \end{split}
 \label{eq:L_lim}
 \end{equation}
where $\mathrm{erf}$ is the Gaussian error function. The likelihoods of the individual elements can then be combined by multiplication:
\begin{equation}
L(x|M) = \prod_{i} L_i(x_i|M).
\label{eq:comb_L}
\end{equation}
The same approach to compute fit likelihoods was also used in, e.g., \citet{Fraser17}. In cases where there are only detections and no upper limits, maximizing this likelihood is equivalent to minimizing $\chi^2$. This way of combining likelihoods implicitly assumes that the errors of all abundance determinations are uncorrelated. Especially for errors from uncertainties in the determination of stellar parameters, this may not be true \citep{McWilliam95}. This is because all low-excitation lines arising from neutral minority species tend to have similar sensitivity to the effective temperature, which typically dominates the error budget. However, we only aim at showing the importance of constraining the dilution of SN ejecta, and a complete treatment of the error distributions and dependencies of abundance determinations exceeds the scope of the current investigation.

If we assign each model in the SN library the same prior probability, we can further compute the probability of each model $M$ given the observations $x$ by
\begin{equation}
 P(M|x) = N \,L(x|M),
\end{equation}
where $N$ is a normalization constant chosen such that
\begin{equation}
 \sum_M P(M|x) = 1.
\end{equation}
 
 \section{Application to observations}
  In this section we demonstrate in three cases why it is important to consider the dilution when fitting abundances of metal-poor stars. Firstly, we will show that it can help to break degeneracies in a fit; secondly, that it may systematically change properties of large fitted samples of stars; and thirdly, that for some stars there may not be a viable single-progenitor scenario to explain the observed abundance patterns.
 
 \subsection{Example 1: The progenitor of HE~0020-1741}
 \label{sect:deg}
 To investigate the impact of the minimum dilution mass on abundance fitting, we firstly fit the CEMP-no star HE~0020-1741 ([Fe/H] = -3.6). \citet{Hansen19} have determined abundances for 13 elements (C, N, O, Mg, Ca, Sc, Ti, Cr, Mn, Ni, Fe, Sr, Ba). As the yields from \citet{HegerWoosley2010} do not include \textit{r-} and $s-$process elements, Sr and Ba are excluded from the fits. Because it is generally underpredicted in the models Sc is treated as an upper limit.

\begin{figure}
 \includegraphics[width=\linewidth]{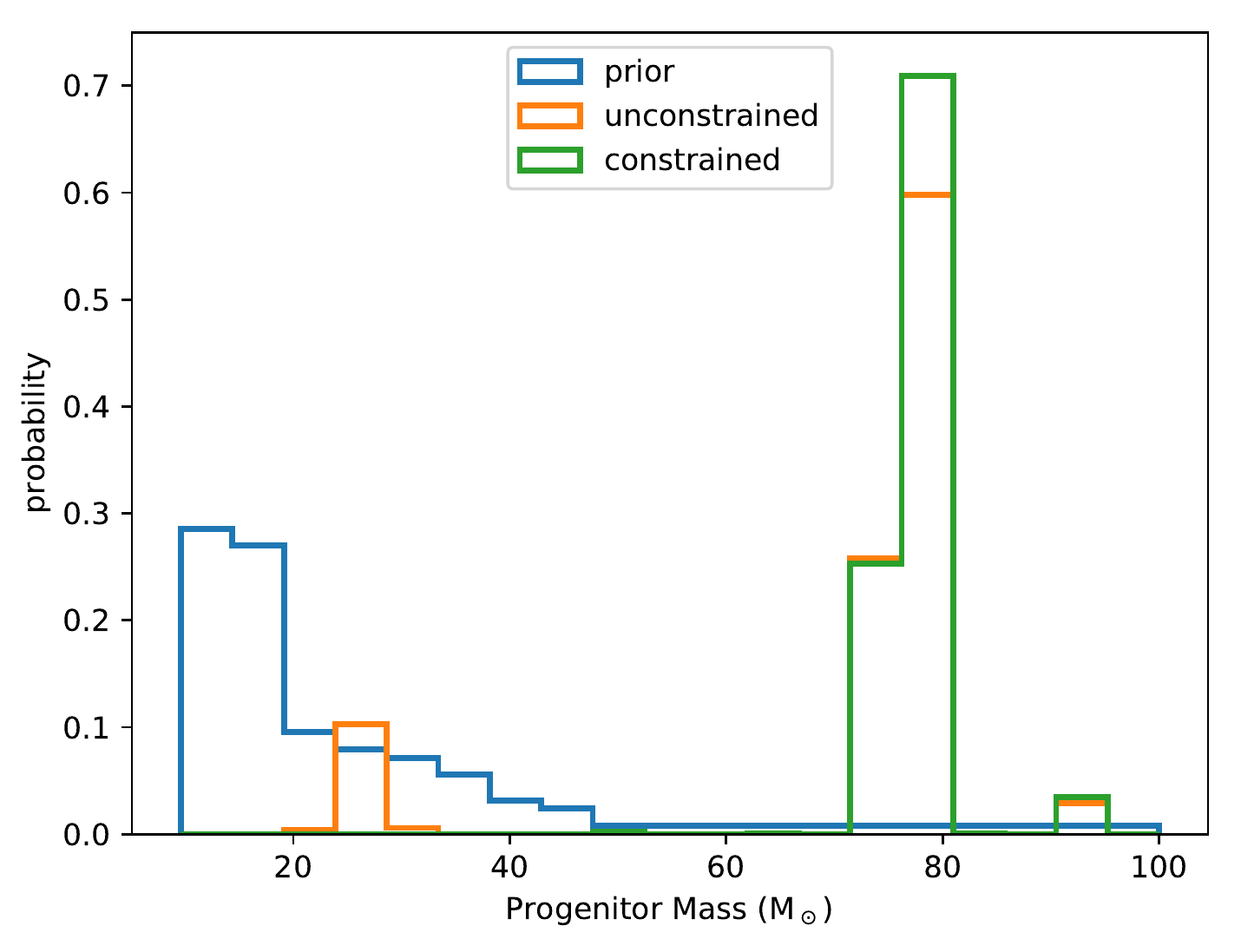}
 \caption{\label{fig:post_M} Prior (blue) and posterior distribution of the progenitor mass of HE~0020-1741. We show the posterior for unconstrained (orange) and constrained (green) dilution. In the unconstrained dilution there is a bimodal distribution of progenitor masses, whereas with the constrained dilution only high-mass stars match the observed abundances.}
\end{figure}

\begin{figure}
 \includegraphics[width=\linewidth]{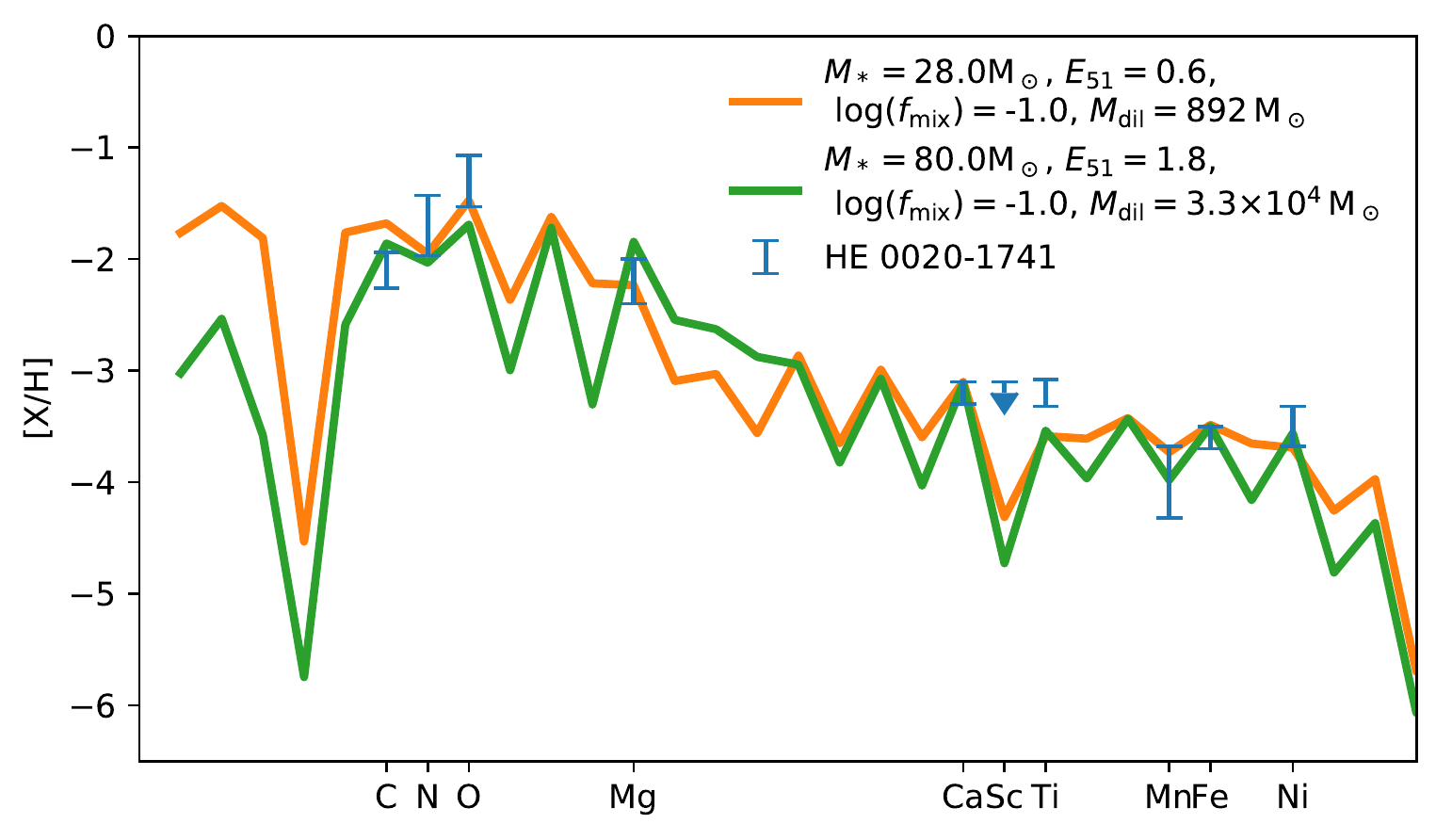}
 \caption{\label{fig:HE-bestfit} Best-fitting abundance patterns for HE~0020-1741. We show the observed pattern and the patterns of the best-fitting models with low-mass and high-mass progenitors. If only the abundance ratios are considered, both give an equally plausible fit, yet constraining the dilution rules out a single low-mass star as a progenitor.}
\end{figure}

We show the prior and the posterior distribution of stellar masses in Fig. \ref{fig:post_M}. The prior is bottom heavy, as there are many more models of low-mass SNe in the libraries than there are models of high-mass SNe. This could potentially bias fitting results towards lower masses. We perform the fits with unconstrained and with constrained dilution factors. In the former case, we chose the dilution factors to maximize the combined likelihoods as defined in Eq. \eqref{eq:comb_L}. In the latter case, we only allow dilution factors above our analytical limit. For the unconstrained dilution we find a bimodal posterior: there are solutions with progenitor masses either around $M_*\approx25\,\Ms$ or at $M_*\approx80\,\Ms$, which we will refer to as ``low-mass'' and ``high-mass'' here. In Fig.~\ref{fig:HE-bestfit} we show the abundance pattern produced by the best-fitting model from each of these branches. Both models seem to fit approximately equally well. However, if we consider only the constrained dilution case, only the high-mass progenitors still fit. The low-mass progenitors do not produce enough metals to explain the metal abundances of HE~0020-1741 with only a single SN. Thus, if HE~0020-1741 is to be explained with a single progenitor SN, it should be a massive star with $70\,\Ms<M_* < 80\,\Ms$  for the \citet{HegerWoosley2010} yields. Note, however, that it may be possible to find additional or better fits with different yield sets \citep[e.g.][]{Limongi12, Ishigaki18, Grimmett18}. However, as the aim here is to show the usefulness of the dilution limit to constrain fits, a comparison of these different yield set exceeds the scope of this study.

 \subsection{Example 2: Large sample fitting}
 \label{sect:ex2}
\begin{figure}
  \includegraphics[width=\linewidth]{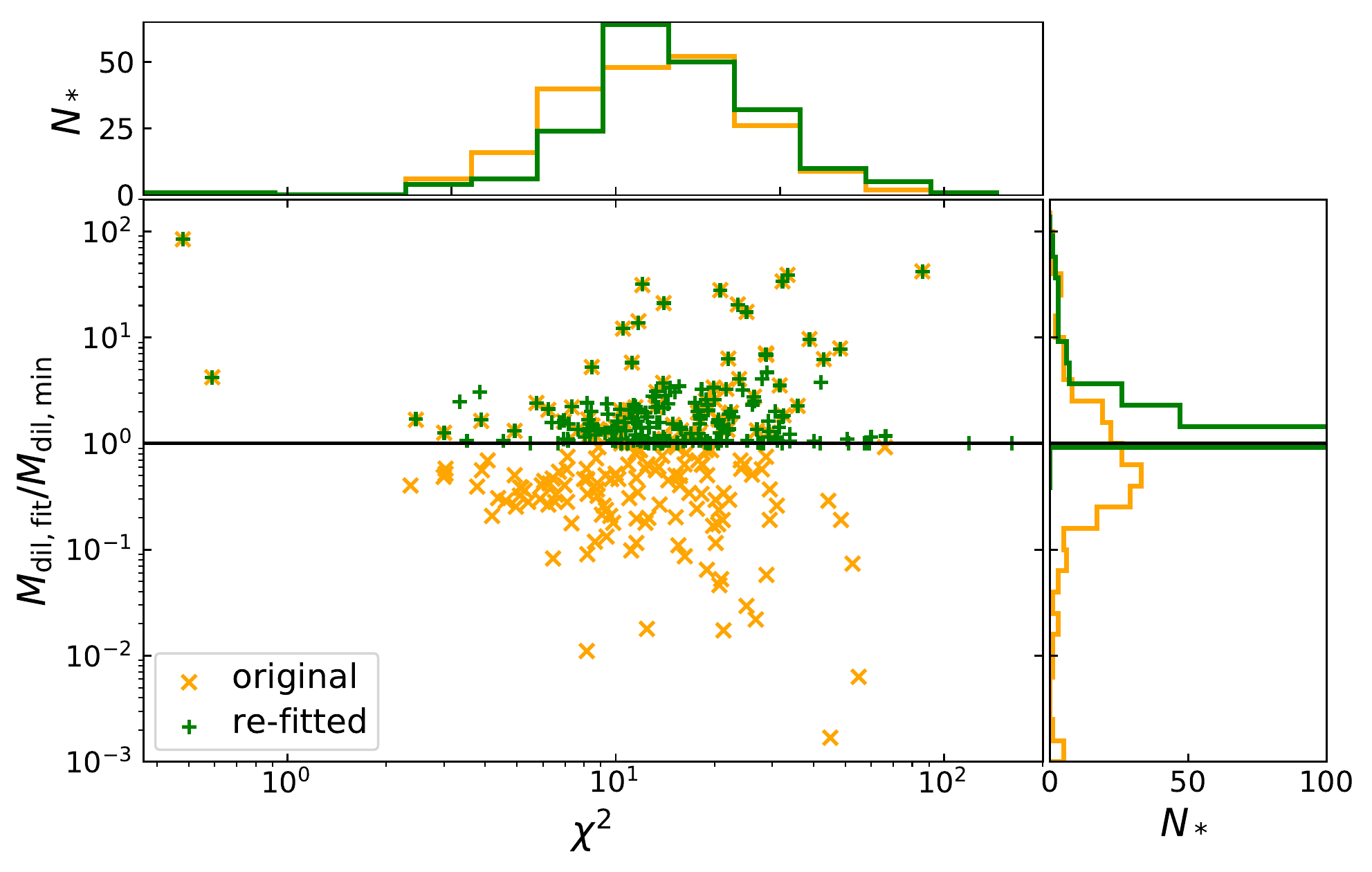}
  \caption{\label{fig:comparison} Ratio of the minimum dilution mass and the dilution mass derived in \citet{Ishigaki18} as function of the reduced $\chi^2$, as well as histograms of both values. We show the original fits from \citet{Ishigaki18} (orange) as well as re-fits in which the minimum dilution limit is enforced (green). In some of the original fits, the dilution ratio is very low (down to $\sim 10^{-7}$) and therefore outside of the boundaries of this figure. These stars are included in the lowest bin of the histogram.}
 \end{figure}

 \citet{Ishigaki18} fitted the abundances of 201 EMP stars by picking the best-fitting SN model for each of these stars. The compiled sample of stars has been selected to consist only of stars with determined abundances for the elements C, N, O, Na, Mg, Al, Si, Ca, Sc, Ti, Cr, Mn, Fe, Co, Ni, and Zn based on spectroscopic data with a resolution of at least $R=28000$. These observed abundances were compared to SN models which were computed over a grid of stellar masses, explosion energies as well as three parameters that quantify the properties of the mixing-and-fallback process. Details on the sample selection and the SN modelling can be found in \citet{Ishigaki18}. Each star in the sample was associated with a best-fitting model, based on $\chi^2$ minimization. In this paper, we use the explosion energies of their best-fit models to compute minimum dilution masses for each of the SN models. In Fig. \ref{fig:comparison} we show the ratio between these minimum dilution masses and the dilution masses from the fits for all 201 stars. Of these 201 best-fitting models, 128 violate our derived limit and 43 do so by more than a factor of four. Notably there is no apparent correlation between $\chi^2$ and the dilution mass. Thus, whether the dilution factor found by fitting is physical is unrelated to the goodness of fit.

\begin{figure}
  \includegraphics[width=\linewidth]{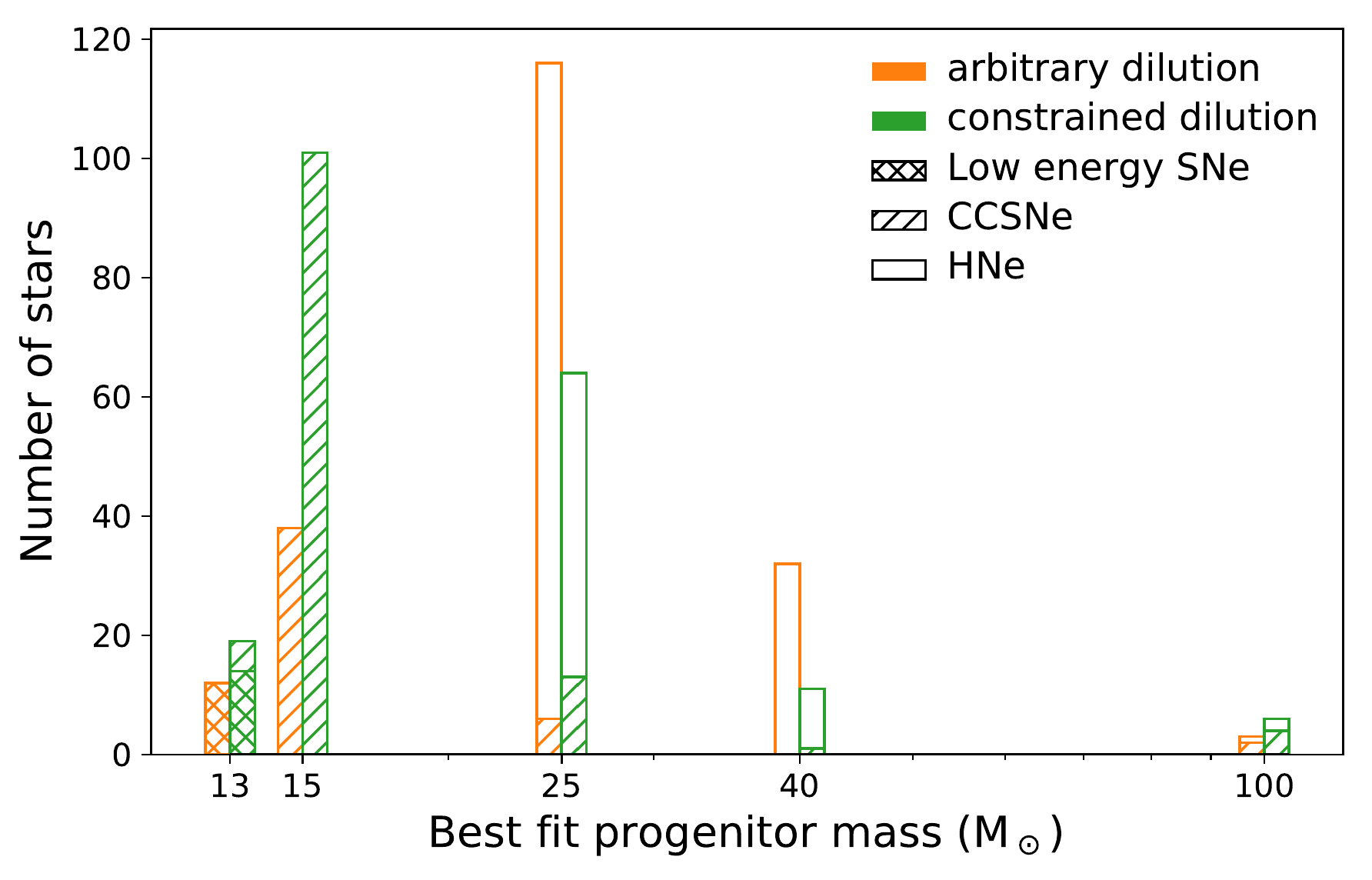}
  \caption{\label{fig:prog_mass_hist} Distribution of progenitor masses from \citet{Ishigaki18} as well as our repetition of the fits which take the minimum dilution criterion into account. There are only bins at 13, 15, 25, 40 and 100\,\Ms\ as these are the only progenitor masses of the SN models. We separate SNe by explosion energy into low-energy SNe ($E_{51}<1$), CCSNe ($E_{51}=1$) and hypernovae (HNe, $E_{51} \ge 10$).}
 \end{figure}

 For comparison, we have repeated the fitting procedure from \citet{Ishigaki18} while at the same time enforcing the dilution limit. This means that all fits with too small dilution masses are rejected and instead the best-fitting model that fulfils the dilution limit is picked. This leads to a significant increase in the mean (median) $\chi^2$ from 16 (13) in the unconstrained case to 24 (15) in the constrained case. For many stars, the best fit with the dilution constraint becomes worse than that without it. In Fig. \ref{fig:prog_mass_hist} we show that this re-fitting leads to significant changes in the distribution of best-fitting progenitors masses. The most notable difference is that progenitors with a stellar mass of 25 and 40\,\Ms\ are now much rarer and progenitors with 15\,\Ms\ more common. The reason for this is that many of the previously common 25 and 40\,\Ms\ models were hypernovae (HNe) with a high explosion energy and a large fallback fraction. These stars have relatively low absolute yields, but due to their large explosion energies, we predict large dilution masses in spherical symmetry. Therefore, such models are not able to reproduce the relatively large carbon abundances of many CEMP-no stars, when taking the dilution constraint into account.
 
We note that the prescription of faint SNe used in \citet{Ishigaki18} is chosen to reproduce the angle-averaged yields of aspherical jet SNe \citep{Tominaga09}. Our dilution model, however, does not apply to such SNe if their asphericity is preserved.  In the used prescription only the total yields are considered. Even if the abundance distribution in the ejecta is strongly aspherical, this approximation assumes that the SN yields are mixed and the angular variations are washed out during later phases of the expansion of the SN. In principle, the mixing behaviour in aspherical SNe  can be very different from our approximation if the metal yield per unit energy shows strong angular variations. Additionally, aspherical SNe from \citet{Ishigaki18} have systematically larger and can have much larger explosion energies, which are used in Equation \eqref{eq:M_dil}, than their 2D counterparts with similar yields \citep{Tominaga09}. This further limits the applicability of our model to these SNe. Our results here suggest that developing realistic models for the dilution of heavy elements produced in aspherical SNe is of vital importance for fitting large samples of stars, not just individual cases.

 \subsection{Example 3: No spherical progenitor for SMSS0313-6708}
 As we realized previously that stars with high carbon and low iron abundances are particularly strongly affected by applying the dilution criterion, we will look in more detail at a pathological example of such a star, i.e., SMSS0313-6708 \citep{Keller14}. The star is known for having no detected iron abundance with an upper limit of $\mbox{[Fe/H]}<-7.1$. We here use abundances that are based on 3D atmospheric models that do not assume local thermodynamical equilibrium (3D, NLTE) for Na, Mg, Al, Ca, and Fe from \citet{Nordlander17}. For these elements statistical and systematic errors are provided which we add with a quadratic sum. The systematic errors are typically on a level of 0.1~dex. The remaining abundances are taken from \citet{Bessel15} and are based on 3D LTE models for C, N, and O and on 1D LTE models for Si, Sc, Ti, V, Cr, Mn, Co, Ni, and Cu. Notably, most of these elements are not detected and only upper limits on their abundance have been derived. \citet{Bessel15} only give statistical but no systematical errors for their abundance determinations. Because we want avoid biasing our results towards abundances with an unaccounted source of error, we add a systematical error of 0.1~dex to the abundance determinations from \citet{Bessel15}.
 
 \begin{figure}
  \includegraphics[width=\linewidth]{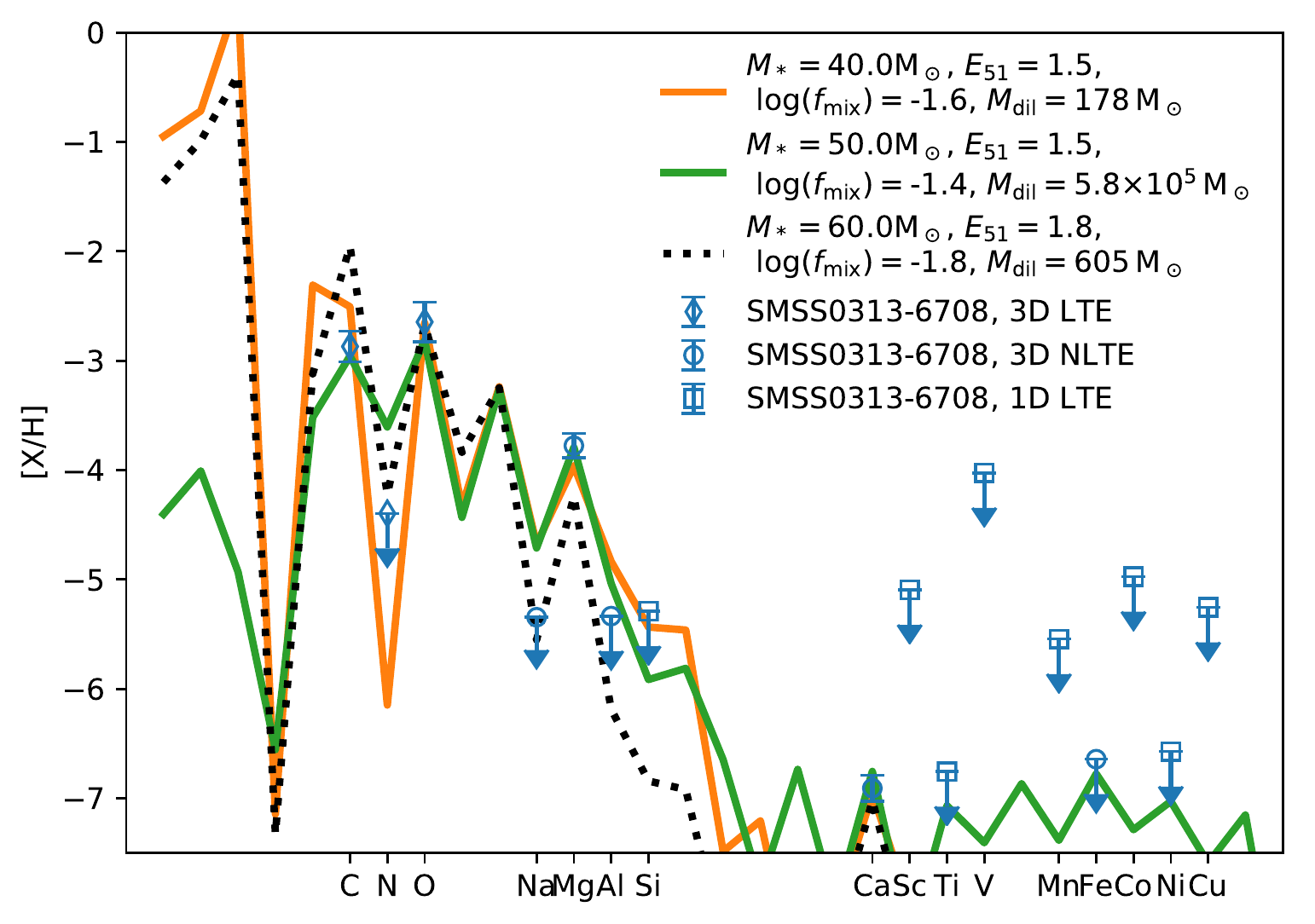}
  \caption{\label{fig:Keller_fit}Best-fitting models for SMSS0313-6708. We show the unconstrained (orange) and the constrained (green) best-fitting model. We find the same best-fitting model as \citet{Bessel15}. For reference we also show the best-fitting model from \citet{Keller14}. We mark with different symbols whether abundances have been derived in 1D LTE (diamonds), 3D LTE (squares), or in 3D NLTE (circles)}  For better representation we shifted the upper limits by one standard deviation, such that the upper end of the error bar corresponds to discrepancy at a 84 per cent confidence level.
 \end{figure}
 
 We fit the abundances with the same procedure as described in Sect. \ref{sect:deg}. The best constrained and unconstrained models are shown in Fig. \ref{fig:Keller_fit}. For better graphical representation we shift the upper limits by one standard deviation. Thus, according to Eq.~\eqref{eq:L_lim} the upper end of the error bar represents a discrepancy at a 84 per cent confidence level, and a value $1\sigma$ above the upper limit corresponds to a 98 per cent significant discrepancy. Even with unconstrained dilution, we find no model that produces a convincing fit of the abundance patterns. The best-fitting model overproduces Na, C and Si. There are three features in the abundances that are difficult to fit simultaneously:
 \begin{enumerate}
  \item the CNO pattern with high C and O but very low N,
  \item the low upper limit on Na with the detection of a large amount of Mg, and
  \item the detection of Ca in conjunction with the low upper limits on Al and Si 
 \end{enumerate}

 The difficulty involved in reproducing all three of these features may partially be related to the grid of models not containing a sufficiently large variety of SN explosion energies. We note that none of the elemental abundances that have been derived in 1D LTE play a critical role in constraining the models. All 1D LTE abundances are only upper limits that lie well above the best-fitting models. It is still unclear how different the C and O abundances would be in a 3D NLTE analysis, but they would need to differ by approximately 1 dex from 3D LTE in order for us to be able to find matching abundance SNe. \citet{Nordlander17} were able to fit the abundance patterns by interpolating the abundance patterns as function of the explosion energy. However, the result of such a procedure is potentially sensitive to the way the interpolation is done. We therefore decided against interpolating to a finer grid here.
 
 Both the unconstrained best-fitting model and the best-fitting model from \citet{Keller14} violate the dilution limit by around two orders of magnitude.  They would require the SN ejecta to be diluted with less than 500\,\Ms\ of pristine material. The best-fitting model that fulfils the dilution limit is clearly inconsistent with the observed upper limits of N and Na.  In \citet{Ishigaki14} this star was best-fitted with $25 M_\odot$ and $40M_\odot$ SN or HNe (jet-induced, aspherical, and energetic SN), where Ca is produce by static/explosive O burning and incomplete Si burning in contrast to the explanation in \citet{Keller14}. Of these models, the SNe are consistent with our dilution limit and the HNe are neither compatible with the dilution criterion nor with the updated upper limit on Si that we use. The fits presented in \citet{Ishigaki18} are compatible both with the abundance pattern we use and with our dilution limit.
\citet{Chen17} model potential progenitor SNe for SMSS0313-6708 in one and two dimensions.  Of their models only the two dimensional model of the SN of a 60\,\Ms\ Pop III star is consistent with our dilution limit.  As these elements are not modelled, however, it is unclear whether this model is able to reproduce the observed upper limits of the  Na and Al abundances. \citet{Chan20} performed full 3D SNe simulations of the progenitor of SMSS0313-6708 using a 40\,\Ms\ star, as suggested by \citet{Bessel15}, with asymmetric explosion of low and high energy.  Their nucleosynthesis has the same constraints as those by \citet{Chen17}, and the explosion was not followed beyond shock breakout. The low-energy model does not produce any significant metals, the high-energy model too much iron -- if spherically averaged.
 
\section{DISCUSSION AND SUMMARY}
We have introduced a analytical limit for the dilution of metals produced by a single SN with the following three assumptions: 
\begin{itemize}
 \item the SN being alone and isolated,
 \item the explosions being spherical and well-mixed, and
 \item the surrounding medium being homogeneous.
\end{itemize}
 The first two assumptions are commonly made when comparing observed abundance patterns to SN yields in previous works, because if these are not fulfilled the total elemental yields from a single SN cannot be representative stellar abundance pattern.  For the last assumption we compared this limit to all hydrodynamical simulations of metal enrichment in high-redshift minihaloes which we are aware of and which included the needed details and resolution for a comparison. We found that, despite assuming homogeneity, the limit is consistent with all of these simulations. 
 
We demonstrate that previous fits were often inconsistent with our understanding of metal dilution and mixing on the scale of minihaloes. Including our dilution criterion into fitting procedures for abundance patterns can have important consequences for the conclusions drawn:
 \begin{enumerate}
  \item Considering the dilution can help to break degeneracies in progenitor models of individual stars.
  \item The limit does not just affect individual stars but it can also change the properties of large samples of progenitor models. In particular, low-yield SNe are disfavoured if constraints on the dilution are taken into account.
  \item It may be difficult to explain certain stars, such as SMSS0313-6708 by enrichment from a single, spherical SN if the dilution is taken into account. The best-fitting models that have been put forward by \citet{Keller14} and \citet{Bessel15} explain the rough shape of the observed abundance ratios, but with an implicit dilution mass that is too small by approximately two orders of magnitude, the yield from the SNe are too small to explain the absolute metal abundances. \citet{Ishigaki14, Ishigaki18} find fits to the abundance pattern that are consistent with our dilution criterion.
 \end{enumerate}
 
During the preparation of this manuscript, \citet{Komiya20} derived a similar estimate for the minimal dilution and implemented it into a semi-analytical model of the formation of the Milky Way. While we apply this estimate to the exploration of progenitor scenarios of individual stars, \citet{Komiya20} focus on the chemical evolution of the Milky Way and in particular on whether the overall population of CEMP-no stars can be reproduced. They find it difficult to reproduce the prevalence of large carbon abundances in the lowest metallicity stars with faint SNe. This tension between the mixing-and-fallback SN model and the large observed carbon abundances is consistent with our findings.
 
 The minimum dilution estimate can serve for evaluating whether a single, spherical SN is a viable progenitor scenario for a certain star. This test may be less reliable, if applicable at all, for asymmetric SNe or cases with several SNe in one halo. In asymmetric SNe, a large fraction of the metals can be ejected along jets \citep{Tominaga09}. Evidence for such SNe has recently been found by \citet{Ezzeddine19}. The dilution and recollapse occuring after such SNe are yet to be explored by numerical simulations.

 Altogether, we conclude that for the adequate astrophysical interpretation of the observed elemental abundances in extremely metal-poor stars, both the relative abundance patterns as well as the absolute abundance values need to be taken into account. Only then reliable and well founded constraints on the properties of the preceding generation of stars can be derived. Given the fact that simple spherically symmetric models often fail to match the dilutions mass constraint introduced here, we furthermore conclude that the effects of aspherical supernovae, the impact of inhomogeneous mixing in a highly structured interstellar medium, and the combined yields of multiple supernovae requires further investigation.

\section*{Acknowledgements}
The authors would like to thank Nozomu Tominaga for very productive discussions and comments. In preparation of this manuscript, the software packages \textsc{F2PY} \citep{f2py}, \textsc{NumPy} \citep{numpy}, \textsc{matplotlib} \citep{matplotlib} and \textsc{SciPy} \citep{SciPy} were used.

MM was supported by the Max-Planck-Gesellschaft via the fellowship of the International Max Planck Research School for Astronomy and Cosmic Physics at the University of Heidelberg (IMPRS-HD). SCOG and RSK acknowledge funding from the Deutsche Forschungsgemeinschaft (DFG) -- Project-ID 138713538 -- SFB 881 (``The Milky Way System'', sub-projects A1, B1, B2 and B8). Further financial support was provided by the DFG via the Heidelberg Cluster of Excellence {\em STRUCTURES} in the framework of Germany’s Excellence Strategy (grant EXC-2181/1 - 390900948).
AH was supported in part by the National Science Foundation under Grant No.\ PHY-1430152 (JINA Center for the Evolution of the Elements), by the the Australian Research Council Centre of Excellence for Gravitational Wave Discovery (OzGrav), through project number CE170100004, by the Australian Research Council Centre of Excellence for All Sky Astrophysics in 3 Dimensions (ASTRO 3D), through project number CE170100013, and by a grant from Science and Technology Commission of Shanghai Municipality (Grants No.16DZ2260200) and National Natural Science Foundation of China (Grants No.11655002).  C.K. acknowledges funding from the UK Science and Technology Facility Council (STFC) through grants ST/M000958/1 and ST/R000905/1. KN has been supported by the World Premier International Research Center Initiative (WPI Initiative), MEXT, Japan, and JSPS KAKENHI Grant Number JP17K05382 and JP20K04024.
 
\section*{Data Availability}
No new data were generated in support of this research. The SN models from \citet{HegerWoosley2010} are available on \url{http://2sn.org}. For the availability SN models or the sample of stars used in Section~\ref{sect:ex2}, please inquire with the authors of \citet{Ishigaki18}.
 

\bibliographystyle{mnras}
\bibliography{lit}

\bsp 
\label{lastpage}
\end{document}